\documentclass[preprint,aps,prd,showpacs]{revtex4}

\usepackage{amsmath}
\usepackage{amsfonts}
\usepackage{graphicx}

\newcommand{\npartial}{\not\!\partial}
\newcommand{\Aslash}{\not\!\!A}
\newcommand{\Pslash}{\not\!\!\,\!\,\,\!\Pi}

\begin{document}

  \preprint{UTEXAS-HEP-01-22}
  \preprint{hep-ph/0106142}
  \title[$Z$ decay in weak magnetic field]{$Z$ decay into two massless
    gauge bosons in a magnetic field}
  \date{August 29, 2001}
  \author{Todd M. Tinsley}
  \email{ttinsley@physics.utexas.edu}
  \affiliation{Center for Particle Physics, University of Texas, Austin,%
              Texas, 78712, USA}
  \begin{abstract}
    An investigation of the processes $Z\rightarrow gg$ and
    $Z\rightarrow \gamma\gamma$ in a background magnetic
    field is presented.  For homogeneous fields corrections to the
    charged fermion propagator can be calculated in leading orders of the magnetic
    field. This work examines the first order contributions of the
    corrected propagator to decays that are otherwise zero.  Results
    of the decay rates for varying field strengths are included.
  \end{abstract}

  \maketitle

  \section{Introduction}
  The interaction of matter with magnetic fields has long been
  studied for the important consequences it has for particle
  dynamics.  Whether it is from the astrophysical implication of a
  magnetic field on the structure of a neutron star or to the Zeeman
  shift of energy levels within an atom, the presence of an
  external magnetic field can open the door to interesting effects
  not previously seen.

  The effect of an homogeneous background field on the charged
  fermion propagator in QED was first studied by Schwinger
  \cite{Schwing}.
  Schwinger's proper-time method provides the machinery for an
  exact solution to the altered propagator.  Recently, a weak field
  expansion of this method has been applied to demonstrate that
  there exists enhancements to neutrino-photon interactions that
  involve charged fermion loops \cite{WeakExp2,DicusNeuPhot,WeakExp}.  The
  goal of this article is to present what effects a magnetic field has
  on the decays of the $Z$ which are otherwise forbidden.

  The Landau-Yang theorem uses Bose symmetry and rotational invariance
  arguments to prove that a vector particle cannot decay into two
  massless vector particles \cite{Landau,Yang}.  That is, the decays
  $Z\rightarrow\gamma\gamma$ and $Z\rightarrow gg$ are not
  allowed.  However we can employ the field expansion of Schwinger's
  method to calculate if and to what degree the magnetic field stimulates
  the decay.

  We expand just to first order in the magnetic field $B$.  This
  turns out to be a very good approximation considering that the
  relevant expansion parameter is $|Qe|B/m^{2}$, where $m$ is the
  mass of the fermion and $Qe$ is its charge.  For the lightest of
  the possible fermions, the electron, the critical field turns out
  to be of the order $B_{0}=m_{e}^{2}/e \approx 4.4\cdot 10^{9}\,\mathrm{T}$.
  This field is a great deal larger than those magnetic fields
  produced by very dense white dwarfs, and near the neutron star
  magnetic field intensity.  The calculation is complete to first
  order, but because of this expansion parameter it is shown
  that the dominant contributions are from the lightest quarks
  ($u,d$) for $Z\rightarrow gg$ and from the electron in the
  $Z \rightarrow\gamma\gamma$ case.

  First we briefly introduce the reader to the methodology used in
  determining the changes in the fermion propagator.  Next, the
  result of that technique
  is presented.  This result is then applied to the
  decay mode of the $Z$ involving two gluons.  The results of this
  calculation are presented and contrasted with results from
  the similar decay into two photons.

  \section{Overview of Schwinger's Result}
  Schwinger's method uses an integration over a proper-time
  variable to solve for the Green's function of the
  inhomogenous differential equation for a Dirac field in the
  presence of the electromagnetic gauge field $A_{\mu}$
  \begin{equation}\label{eqn:Dirac}
    (i\!\!\npartial-eQ\!\!\Aslash-m)G(x,x^{\prime})=\delta^{4}(x-x^{\prime}).
  \end{equation}
  $G(x,x^{\prime})$ is considered the matrix element of an
  operator $G$, such that
  $G(x,x^{\prime})=\langle x|G|x^{\prime}\rangle$.  Therefore
  Eq.\ (\ref{eqn:Dirac}) is written as an operator equation
  with its indices suppressed
  \begin{equation}\label{eqn:DiracMatrix}
    (\Pslash-m)G=1,
  \end{equation}
  or
  \begin{eqnarray}\label{eqn:Gproptime}
    G &=& \frac{1}{\Pslash-m}=%
          \frac{\Pslash+m}{\Pslash^{2}-m^{2}}\nonumber\\
      &=& -i\int_{0}^{\infty}{\mathrm{d}}s%
          (\Pslash+m)\exp{[i(\Pslash^{2}-m^{2})s]}.
  \end{eqnarray}
  $s$ is the proper-time variable, and $\Pi_{\mu}$ is a generalized
  momentum operator defined by
  \begin{equation}\label{eqn:GenMom}
    \Pi_{\mu}=p_{\mu}-eQA_{\mu}.
  \end{equation}
  This generalized momentum operator is characterized by the
  following commutation properties
  \begin{subequations}\label{eqn:Commutators}
  \begin{eqnarray}
    [\Pi_{\mu},x_{\nu}]&=&ig_{\mu\nu}\\\label{eqn:Com1}
    [\Pi_{\mu},\Pi_{\nu}]&=&-ieQF_{\mu\nu},\label{eqn:Com2}
  \end{eqnarray}
  \end{subequations}
  where $F_{\mu\nu}$ is the electromagnetic field-strength tensor.

  The quantity $\exp[i\!\!\Pslash^{2}s]$ is regarded as an operator that
  evolves the system in accordance with the ``Hamiltonian''
  ${\mathcal{H}}=-\Pslash^{2}$.  That is, the matrix element of the
  operator $U(s)=\exp[i\!\!\Pslash^{2}s]$ is the time development
  function that takes the state $x_{\mu}(0)=x_{\mu}^{\prime}$ to
  the state $x_{\mu}(s)=x_{\mu}$
  \begin{equation}\label{eqn:UElement}
    \langle x|U(s)|x^{\prime}\rangle = \langle
    x(s)|x(0)\rangle.
  \end{equation}

  This view of the transformation as an evolution of the state
  governed by a Hamiltonian parallels the idea that the fermion
  moves from one space-time location to another in a way that is
  determined by this Hamiltonian.  Accordingly, there exist
  ``equations of motion'' satisfied by the generalized position
  and momentum operators
  \begin{subequations}\label{eqn:EqnOfMot}
  \begin{eqnarray}
    \frac{{\mathrm{d}}x_{\mu}}{{\mathrm{d}}s} &=&
      -i[x_{\mu},{\mathcal{H}}]=2\Pi_{\mu}\\\label{eqn:EqnOfMota}
    \frac{{\mathrm{d}}\Pi_{\mu}}{{\mathrm{d}}s} &=&
      -i[\Pi_{\mu},{\mathcal{H}}]=2QeF_{\mu\nu}\Pi^{\nu}.\label{eqn:EqnOfMotb}
  \end{eqnarray}
  \end{subequations}
  The relationship in Eq.\ (\ref{eqn:EqnOfMotb}) only
  holds when $F_{\mu\nu}$ is a constant.  This is precisely the
  case in which we are concerned.  For calculation purposes the
  magnetic field is considered to be constant and directed along
  the \emph{z}-axis ($F_{12}=-F_{21}=B$, and $F_{ij}=0$ for
  $i,j\neq 1,2$).  In addition there are three differential
  equations that describe the transformation function
  \begin{subequations}\label{eqn:TranDiffEq}
  \begin{eqnarray}
    i\partial_{s}\langle x(s)|x(0)\rangle%
      &=& \langle x(s)|{\mathcal{H}}|x(0)\rangle\\\label{eqn:TranDiffEqa}
    (i\partial_{\mu}+eQA_{\mu})\langle x(s)|x(0)\rangle%
      &=& \langle x(s)|\Pi_{\mu}(s)|x(0)\rangle\\\label{eqn:TranDiffEqb}
    (-i\partial_{\mu}^{\prime}+%
      eQA_{\mu})\langle x(s)|x(0)\rangle%
      &=&\langle x(s)|\Pi_{\mu}(0)|x(0)\rangle,\label{eqn:TranDiffEqc}
  \end{eqnarray}
  \end{subequations}
  and a boundary condition
  \begin{equation}\label{eqn:BoundCond}
    \langle x(s)|x(0)\rangle |_{s\rightarrow 0} = \delta^{4}(x-x^{\prime}).
  \end{equation}

  Using Eqs.\ (\ref{eqn:EqnOfMot}), (\ref{eqn:TranDiffEq}),
  and (\ref{eqn:BoundCond}) one can solve for the explicit form
  of the Green's function.  For the purposes of this article, only the
  result will be given \footnote{For an excellent treatment of
  this solution see Ref. \cite{WeakExp} or Schwinger's own work
  in Ref. \cite{Schwing}.}.  Choosing a straight line to
  connect the space-time points $x$ and $x^{\prime}$, the Green's
  function is
  \begin{equation}\label{eqn:GreenFunct}
    G(x,x^{\prime})=\Phi(x,x^{\prime})\int\frac{{\mathrm{d}}^{4}p}
      {(2\pi)^{4}}e^{-ip(x-x^{\prime})}{\mathcal{G}}(p),
  \end{equation}
  with
  \begin{equation}\label{eqn:Phase}
    \Phi(x,x^{\prime})=\exp{\biggl[-ieQ\int_{x^{\prime}}^{x}
      {\mathrm{d}}\xi^{\mu}A_{\mu}(\xi)\biggr]},
  \end{equation}
  and
  \begin{eqnarray}\label{eqn:GreenFunctMoment}
    {\mathcal{G}}(p)&=&-i\int_{0}^{\infty}
      {\mathrm{d}}s\exp{\Biggl[-is\biggl(m^2-p^{2}_{\parallel}+
      \frac{\tan(eQBs)}{eQBs}p^{2}_{\perp}\biggr)\Biggr]}\nonumber\\
      &&\times\Biggl[e^{ieQBs\sigma_{3}}
      \frac{\bigl(m + \not\!p_{\parallel}\bigr)}{\cos(eQBs)}
      -\frac{\not\!p_{\perp}}{\cos^{2}(eQBs)}\Biggr].
  \end{eqnarray}
  The quantity $p^{2}_{\parallel}$ indicates
  $p^{2}_{0}-p^{2}_{3}$, while
  $p^{2}_{\perp}=p^{2}_{1}+p^{2}_{2}$.  Similarly,
  $\not\!\!p_{\parallel}=\gamma_{0}p_{0}-\gamma_{3}p_{3}$ and
  $\not\!p_{\perp}=\gamma_{1}p_{1}+\gamma_{2}p_{2}$.  By expanding
  Eq.\ (\ref{eqn:GreenFunctMoment}) to first order in the
  magnetic field the integration simplifies, and our result is
  \begin{equation}\label{eqn:GFM1O}
    {\mathcal{G}}(p) \approx (\not\!p-m)^{-1}
      -\frac{ieQ}{2}F_{\alpha\beta}
      \frac{\gamma^{\alpha}\gamma^{\beta}\bigl
      (\not\!p_{\parallel}+m\bigr)}{(p^{2}-m^{2})^{2}}.
  \end{equation}
  By making use of the fact that the magnetic field is
  homogeneous, the result in Eq.\ (\ref{eqn:GFM1O}) was
  cast into covariant notation.

  \section{Application to $Z \rightarrow \lowercase{gg}$}  
\begin{figure}[h]
    \centering \includegraphics[width=40mm]{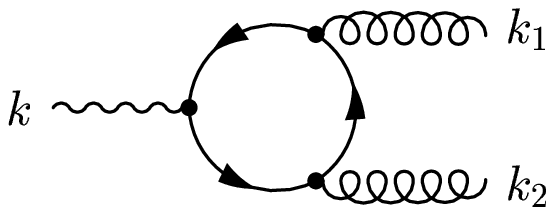}
    \includegraphics[width=40mm]{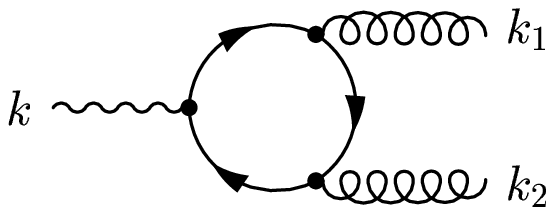}
    \caption{The contributions to the decay of the $Z$ with
    momentum $k$ into two gluons with momenta $k_{1}$ and $k_{2}$.}
    \label{fig:Zggs}
  \end{figure}

At this point
  the general results from the previous section are used to
  calculate the decay rate of the neutral $Z$ boson into to two
  gluons.  To lowest order in the perturbation theory there are
  two Feynman diagrams which contribute to this process; see
  Fig.\ \ref{fig:Zggs}.  Let us
  first turn our attention to the leftmost contribution of
  Fig.\ \ref{fig:Zggs}.%

  The \emph{T-Matrix} for the first of our diagrams is written
  as
  \begin{eqnarray}\label{eqn:Ta}
    T_{A}&=&\sum_{i}\sum_{a,b}\frac{\pi^{3/2}\alpha_{s}\sqrt{\alpha}}
      {\cos{\theta_{W}}\sin{\theta_{W}}}
      {\mathrm{tr}}(\lambda^{a}\lambda^{b})
      \int{\mathrm{d}}^{4}x\int{\mathrm{d}}^{4}y
      \int{\mathrm{d}}^{4}z\nonumber\\
    &&\times{\mathrm{tr}}\bigl[\gamma^{\mu}G^{i}(x,y)\gamma^{\nu}G^{i}(y,z)
      \gamma^{\sigma}(g_{V}^{i}-g_{A}^{i}\gamma^{5})G^{i}(z,x)\bigr]
      \nonumber\\
    &&\times\epsilon_{\mu}^{*a}(k_{1})\epsilon_{\nu}^{*b}(k_{2})
      \epsilon_{\sigma}(k)\exp{[-i(kz-k_{1}x-k_{2}y)]}.
  \end{eqnarray}
  The sum over $i$ indicates the sum over all possible fermions
  that contribute to this loop diagram.  The sums over $a$ and
  $b$ indicate the sums over the possible states of the gluons.
  However, one of these sums is eliminated by virtue of the trace over the
  two SU(3) generators $\mathrm{tr}(\lambda^{a}\lambda^{b})
  =2\delta^{ab}$.  From this point the summation over color will
  be represented by the repeated color index $a$.
  The couplings $g_{V}^{i}$ and $g_{A}^{i}$ are the vector and
  axial vector couplings, respectively; both are dependent on
  the particular fermion that couples to the $Z$ boson.  $\alpha_{s}$ is
  the strong coupling constant evaluated at $M_{Z}$.

  Because there is the combination of the three Green's functions
  in Eq.\ (\ref{eqn:Ta}), there exists a product of three
  phases of the same form as in Eq.\ (\ref{eqn:Phase}).  As
  indicated previously, the integration from two space-time points
  is along a straight line.  A convenient choice of gauge for
  this integration is one in which
  \begin{equation}\label{eqn:Gauge}
    A(\xi)=\frac{B}{2}(0,z_{2}-\xi_{2},\xi_{1}-z_{1},0).
  \end{equation}
  Generally, the zeroth and third components of $A_{\mu}$ can be
  any constant but the above is chosen for its simplicity.  With
  the choice of gauge found in Eq.\ (\ref{eqn:Gauge}) the
  total phase factor of the diagram is written
  \begin{eqnarray}\label{eqn:TotPhase}
     \Phi(x,y)\Phi(y,z)\Phi(z,x)&=&
       \exp{\Bigl[\frac{ieQ}{2}(z-y)^{\mu}
       F_{\mu\nu}(z-x)^{\nu}\Bigr]}\nonumber\\
     &\approx&1+\!\frac{ieQ}{2}(z-y)^{\mu}
       F_{\mu\nu}(z-x)^{\nu}.
  \end{eqnarray}
  The fact that the magnetic field is constant and along the
  \emph{z}-axis has been exploited.  When the first order
  corrections to the Green's function
  found in Eqs.\ (\ref{eqn:GFM1O}) and (\ref{eqn:TotPhase})
  are substituted into Eq.\ (\ref{eqn:Ta}), the result is
  thought of as the sum of five separate contributions
  \begin{equation}\label{eqn:Ta5}
    T_{A} = T_{A0} + T_{A\mathrm{pr}1} + T_{A\mathrm{pr}2} + T_{A\mathrm{pr}3} + T_{A\mathrm{ph}}.
  \end{equation}
  There is the zeroth order contribution to the diagram $T_{A0}$
  which is completely independent of the magnetic field.  There is
  a first order correction to each of the legs of the fermion loop
  $T_{A\mathrm{pr}1}$, $T_{A\mathrm{pr}2}$ and $T_{A\mathrm{pr}3}$.  And lastly, there is a
  first order correction to the overall phase of the diagram
  $T_{A\mathrm{ph}}$.  Because the invariant amplitude of a given diagram
  ${\mathcal{M}}$ is defined through the relationship
  $T=(2\pi)^{4}{\mathcal{M}}\delta^{4}(k-k_{1}-k_{2})$, it is
  straightforward to show that the zeroth order contribution from
  the first diagram is
  \begin{eqnarray}\label{eqn:Ma0}
    {\mathcal{M}}_{A0}&=&\frac{2\pi^{3/2}\alpha_{s}\sqrt{\alpha}}
      {\cos{\theta_{W}}\sin{\theta_{W}}}
      \sum_{i}\sum_{a}\epsilon_{\mu}^{*a}(k_{1})\epsilon_{\nu}^{*a}(k_{2})
      \epsilon_{\sigma}(k)
      \nonumber\\
    &&\times\int\frac{{\mathrm{d}}^{4}p}{(2\pi)^{4}}
      {\mathrm{tr}}\Bigl[\gamma^{\mu}(\not\!p\,-m)^{-1}\gamma^{\nu}
      (\not\!p\,\,+\not\!k_{2}-m)^{-1}\nonumber\\
    &&\times\gamma^{\sigma}(g_{V}^{i}-g_{A}^{i}\gamma^{5})
      (\not\!p\,\,-\not\!k_{1}-m)^{-1}\Bigr].
  \end{eqnarray}
  For the first order correction to one of the legs of
  the loop, the contribution to the amplitude is
  \begin{eqnarray}\label{eqn:Mapr1}
    {\mathcal{M}}_{A\mathrm{pr}1}&=&\frac{-i\pi^{3/2}eF_{\alpha\beta}\alpha_{s}\sqrt{\alpha}}
      {\cos{\theta_{W}}\sin{\theta_{W}}}\sum_{i}\sum_{a}Q
      \epsilon_{\mu}^{*a}(k_{1})\epsilon_{\nu}^{*a}(k_{2})
      \epsilon_{\sigma}(k)\nonumber\\
    &&\times\int\frac{{\mathrm{d}}^{4}p}{(2\pi)^{4}}
      {\mathrm{tr}}\biggl[\gamma^{\mu}
      \frac{\gamma^{\alpha}\gamma^{\beta}(\not\!p\,+m)}
      {(p^{2}-m^{2})^{2}}\gamma^{\nu}
      (\not\!p\,\,+\not\!k_{2}-m)^{-1}\nonumber\\
    &&\times\gamma^{\sigma}(g_{V}^{i}-g_{A}^{i}\gamma^{5})
      (\not\!p\,\,-\not\!k_{1}-m)^{-1}\biggr].
  \end{eqnarray}
  The corrections to the remaining legs of the fermion loop are
  similar to Eq.\ (\ref{eqn:Mapr1}).  However, for the
  contribution due to the phase it is necessary we recognize that
  \begin{eqnarray}\label{eqn:TphPro}
    T_{A\mathrm{ph}} &\propto& (z-y)^{\mu}(z-x)^{\nu}e^{-ip^{\prime}(y-z)}
      e^{-ip^{\prime\prime}(z-x)}\nonumber\\
      &&\times{\mathrm{tr}}\Bigl[\gamma^{\mu}(\not\!p-m)^{-1}\gamma^{\nu}
      (\not\!p^{\prime}-m)^{-1}\gamma^{\sigma}(g_{V}^{i}-g_{A}^{i}\gamma^{5})
      (\not\!p^{\prime\prime}-m)^{-1}\Bigr],
  \end{eqnarray}
  where $p^{\prime}$ is the momentum carried by the fermion as it
  moves from point $z$ to point $y$, and $p^{\prime\prime}$ is its
  momentum from point $x$ to $z$.  The relationship in Eq.\ (\ref{eqn:TphPro})
  is rewritten in terms of derivatives with respect to the fermion
  momenta
  \begin{eqnarray}\label{eqn:TphPro2}
    T_{A\mathrm{ph}} &\propto& \biggl(\frac{\partial}{\partial p^{\prime}_{\mu}}
      \frac{\partial}{\partial p^{\prime\prime}_{\nu}}e^{-ip^{\prime}(y-z)}
      e^{-ip^{\prime\prime}(z-x)}\biggr)\nonumber\\
      &&\times{\mathrm{tr}}\Bigl[\gamma^{\mu}(\not\!p-m)^{-1}\gamma^{\nu}
      (\not\!p^{\prime}-m)^{-1}\gamma^{\sigma}(g_{V}^{i}-g_{A}^{i}\gamma^{5})
      (\not\!p^{\prime\prime}-m)^{-1}\Bigr].
  \end{eqnarray}
  By making use of integration by parts and the identity
  \begin{displaymath}
    \frac{\partial}{\partial x}{\mathcal{C}}^{-1}(x)=-{\mathcal{C}}^{-1}(x)
      \frac{\partial{\mathcal{C}}(x)}{\partial
      x}{\mathcal{C}}^{-1}(x),
  \end{displaymath}
  the first order contribution of the correction to the phase to
  the amplitude is
  \begin{eqnarray}\label{eqn:Maph}
    {\mathcal{M}}_{A\mathrm{ph}}&=&\frac{-i\pi^{3/2}eF_{\alpha\beta}\alpha_{s}\sqrt{\alpha}}
      {\cos{\theta_{W}}\sin{\theta_{W}}}\sum_{i}\sum_{a}Q
      \epsilon_{\mu}^{*a}(k_{1})\epsilon_{\nu}^{*a}(k_{2})
      \epsilon_{\sigma}(k)\nonumber\\
    &&\times\int\frac{{\mathrm{d}}^{4}p}{(2\pi)^{4}}
      {\mathrm{tr}}\bigl[\gamma^{\mu}(\not\!p-m)^{-1}\gamma^{\nu}
      (\not\!p\,\,+\not\!k_{2}-m)^{-1}\gamma^{\beta}
      (\not\!p\,\,+\not\!k_{2}-m)^{-1}\nonumber\\
    &&\times\gamma^{\sigma}(g_{V}^{i}-g_{A}^{i}\gamma^{5})
      (\not\!p\,\,-\not\!k_{1}-m)^{-1}\gamma^{\alpha}
      (\not\!p\,\,-\not\!k_{1}-m)^{-1}\bigr].
  \end{eqnarray}

  We follow the same procedure for the second diagram of Fig.\
  \ref{fig:Zggs} as we did for the first.  The results obtained
  are similar to those found in Eqs.\ (\ref{eqn:Ma0}),
  (\ref{eqn:Mapr1}), and (\ref{eqn:Maph}) with only the labels $(1,2)$
  interchanged.

  \section{The Decay Rate}The sum of the zeroth
  order contributions from the two diagrams vanishes.  This is
  consistent with the fact that the process $Z\rightarrow gg$ is
  forbidden by the Landau-Yang Theorem in the absence of a
  magnetic field \cite{Landau,Yang}.  The use of the charge
  conjugation operator simplifies the contributions that are
  first order in the magnetic field.  We find that the axial
  vector parts of the two diagrams cancel while the vector parts
  are identical and add.  Tracing over the Dirac indices and
  integrating over the loop momentum results in an amplitude that
  is proportional to forty-four separate tensor structures formed
  by possible contractions of the polarization vectors, the
  field-strength tensor, and the gluon momenta.  Since the
  amplitude is symmetric under interchange of the labels $(1,2)$, we
  find that the tensor structures can be grouped into twenty-two pairs;
  each pair having its own unitless coefficient which depends on the possible
  fermion masses, charges, and couplings in the charge loop.  Limiting the
  gluons to be only transversely polarized allows us to drop all terms
  proportional to $(\epsilon_{i}^{*a}\cdot k_{i})$.  This cuts the number
  of tensor structures in half to eleven.  Thus, the amplitude is written as
    \begin{eqnarray}\label{eqn:Amp11}
      {\mathcal{M}}_{Z\rightarrow gg}&=&\biggl(\frac{e\alpha_{s}\sqrt{\alpha}}
        {32\sqrt{\pi}M_{Z}\sin\theta_{W}\cos\theta_{W}}\biggr)
        \times\biggl\{N_{1}\bigl(\epsilon\cdot(k_{1}-k_{2})\bigr)
        (\epsilon_{1}^{*a}F\epsilon_{2}^{*a})(1/M_{Z}) \nonumber \\
       & &+N_{2}\bigl[(\epsilon_{1}^{*a}\cdot\epsilon)(\epsilon_{2}^{*a}F
        k_{1})+(\epsilon_{2}^{*a}\cdot\epsilon)(\epsilon_{1}^{*a}F
        k_{2})\bigr](1/M_{Z}) \nonumber \\
       & &+N_{3}\bigl[(\epsilon_{1}^{*a}\cdot\epsilon)(\epsilon_{2}^{*a}F
        k_{2})+(\epsilon_{2}^{*a}\cdot\epsilon)(\epsilon_{1}^{*a}F
        k_{1})\bigr](1/M_{Z}) \nonumber \\
       & &+N_{4}(\epsilon_{1}^{*a}\cdot\epsilon_{2}^{*a})
        \bigl(\epsilon F (k_{1}+k_{2})\bigr)(1/M_{Z}) \nonumber \\
       & &+N_{5}(\epsilon_{1}^{*a}\cdot\epsilon_{2}^{*a})
        \bigl(\epsilon\cdot(k_{1}-k_{2})\bigr)(k_{1} F k_{2})
        (1/M_{Z})^{3} \nonumber \\
       & &+N_{6}(\epsilon_{1}^{*a}\cdot k_{2})(\epsilon_{2}^{*a}\cdot k_{1})
        \bigl(\epsilon\cdot(k_{1}-k_{2})\bigr)(k_{1} F k_{2})
        (1/M_{Z})^{5} \nonumber \\
       & &+N_{7}\bigl[(\epsilon_{1}^{*a}\cdot\epsilon)
        (\epsilon_{2}^{*a}\cdot k_{1})-(\epsilon_{2}^{*a}\cdot\epsilon)
        (\epsilon_{1}^{*a}\cdot k_{2})\bigr](k_{1} F k_{2})
        (1/M_{Z})^{3} \nonumber \\
       & &+N_{8}\bigl[(\epsilon_{2}^{*a}\cdot k_{1})
        (\epsilon\cdot k_{2})(\epsilon_{1}^{*a}F
        k_{2})+(\epsilon_{1}^{*a}\cdot k_{2})
        (\epsilon\cdot k_{1})(\epsilon_{2}^{*a}F
        k_{1})\bigr](1/M_{Z})^{3} \nonumber \\
       & &+N_{9}\bigl[(\epsilon_{2}^{*a}\cdot k_{1})
        (\epsilon\cdot k_{1})(\epsilon_{1}^{*a}F
        k_{2})+(\epsilon_{1}^{*a}\cdot k_{2})
        (\epsilon\cdot k_{2})(\epsilon_{2}^{*a}F
        k_{1})\bigr](1/M_{Z})^{3} \nonumber \\
       & &+N_{10}\bigl[(\epsilon_{2}^{*a}\cdot k_{1})
        (\epsilon\cdot k_{2})(\epsilon_{1}^{*a}F
        k_{1})+(\epsilon_{1}^{*a}\cdot k_{2})
        (\epsilon\cdot k_{1})(\epsilon_{2}^{*a}F
        k_{2})\bigr](1/M_{Z})^{3} \nonumber \\
       & &+N_{11}(\epsilon_{1}^{*a}\cdot k_{2})
        (\epsilon_{2}^{*a}\cdot k_{1})\bigl(\epsilon F (k_{1}+ k_{2})\bigr)
        (1/M_{Z})^{3}\Bigr\},
    \end{eqnarray}
  where $M_{Z}$ is the mass of the $Z$, $\theta_{W}$ is the Weinberg
  angle, and the $N_{i}$ are the unitless coefficients.  The reader should
  see the appendix for a detailed explanation of how the coefficients
  are defined.

  At this point we confirm that our amplitude is gauge invariant by
  replacing a polarization vector by its associated momentum
    \begin{equation}\label{eqn:gaugeInv}
      \epsilon_{i}^{*\mu}(k_{i})\rightarrow k_{i}^{\mu},
    \end{equation}
  and verifying that the amplitude vanishes.  For these purposes we drop
  the color index $a$.  The gauge invariance implies six independent
  relations among the coefficients that reduce the number
  of independent coefficients from eleven to five.  These six relations and five
  coefficients are given explicitly in Eqs.\,(\ref{eqn:GIRelat}) and (\ref{eqn:Coeff})
  of the appendix.

  What is of real interest, however, is the square of this amplitude.  When the
  square is taken and evaluated in the center of mass frame of the $Z$, we find
  that the square of the amplitude depends on only three of our five
  independent tensor coefficients
  \begin{equation}\label{eqn:Amp2}
    |{\mathcal{M}}_{Z\rightarrow gg}|^{2}=
      \frac{e^{2}\alpha\alpha_{s}^{2}B^{2}}
      {256\pi M_{Z}^{2}\cos^{2}\theta_{W}\sin^{2}\theta_{W}}
      \bigl(4|N_{1}|^{2}\cos^{2}\theta
        +|N_{2}-N_{3}|^{2}\sin^{2}\theta\bigr),
  \end{equation}
  where $\theta$ is the polar angle as measured from the
  direction in which the magnetic field points.

  It is useful to extract from Eq.\ (\ref{eqn:Amp2}) the
  effective expansion parameter for the magnetic field.  That is, it was
  discussed earlier that the perturbation theory for the magnetic
  field is an expansion in $|Qe|B/m^{2}$.
  Because this is fermion dependent, it is useful to pick the
  lightest fermion and make $B/B_{0}$ the expansion parameter.
  Recall that $B_{0}=m_{e}^{2}/e$.  Making this
  substitution, Eq.\ (\ref{eqn:Amp2}) becomes
  \begin{equation}\label{eqn:Amp2a}
    |{\mathcal{M}}_{Z\rightarrow gg}|^{2}=
      \frac{\alpha\alpha_{s}^{2}M_{Z}^{2}(B/B_{0})^{2}}
      {256\pi\cos^{2}\theta_{W}\sin^{2}\theta_{W}}
      \biggl(\frac{m_{e}}{M_{Z}}\biggr)^{4}
      \bigl(4|N_{1}|^{2}\cos^{2}\theta
        +|N_{2}-N_{3}|^{2}\sin^{2}\theta\bigr).
  \end{equation}

  At this point it is instructive to note that the leading
  contributions of $N_{1}$ go as $\ln\bigl(\frac{M_{Z}}{m}\bigr)^{2}$,
  while for both $N_{2}$ and $N_{3}$ the leading contributions go
  as $\bigl(\frac{M_{Z}}{m}\bigr)^{2}$ \footnote{One can obtain this
  result by expanding the relationships in Eqs.\ (\ref{eqn:N1}),
  (\ref{eqn:N2}), (\ref{eqn:N3}), (\ref{eqn:At}), and
  (\ref{eqn:Bt}) of the appendix for large values of
  $\bigl(\frac{M_{Z}}{m}\bigr)^{2}$.}.
  Qualitatively, this means that only the lightest of the fermions
  make significant contributions to the the decay rate.  It also
  means that the $Z$ is many times more likely to decay such that
  the gluons travel perpendicular to the magnetic field rather
  than parallel.  Computation reveals that this factor is
  \begin{displaymath}
    2\cdot 10^{13}\gtrsim\frac{|N_{2}-N_{3}|^{2}}{4|N_{1}|^{2}}\gtrsim 6\cdot
    10^{10}.
  \end{displaymath}
  The large range in values is a symptom of the very large uncertainty
  in the current quark masses.  If the constituent quark masses are used,
  the ratio drops to a factor of $2\cdot 10^{4}$.

  The decay rate for this process is
  \begin{eqnarray}\label{eqn:Drate}
    \Gamma_{Z\rightarrow gg} &=&
      \frac{\alpha\alpha_{s}^{2}M_{Z}(B/B_{0})^{2}}{3\cdot
      2^{11}\pi^{2}\cos^{2}\theta_{W}\sin^{2}\theta{w}}
      \biggl(\frac{m_{e}}{M_{Z}}\biggr)^{4}
      \bigl(2|N_{1}|^{2}+|N_{2}-N_{3}|^{2}\bigr).
  \end{eqnarray}
  Because of the large current quark mass ranges quoted by the Particle
  Data Group \cite{PDBook}, there is a corresponding large range in
  possible values for the decay rate.  We choose to use the mean
  values of these masses for computation.  Seperate calculations are
  carried out using the constituent quark masses.
  Given below are the results of the decay rate for both the current
  and constituent mass calculations \footnote{The current quark masses
  that are used in this calculation are: $m_{u}=3\,{\mathrm{MeV}}$,
  $m_{d}=6\,{\mathrm{MeV}}$, $m_{s}=122.5\,{\mathrm{MeV}}$,
  $m_{c}=1.25\,{\mathrm{GeV}}$, $m_{b}=4.2\,{\mathrm{GeV}}$, and
  $m_{t}=174.3\,{\mathrm{GeV}}$.  The constituent quark mass values used are:
  $m_{u}=m_{d}=350\,{\mathrm{MeV}}$, $m_{s}=550\,{\mathrm{MeV}}$,
  $m_{c}=1.8\,{\mathrm{GeV}}$, $m_{b}=4.5\,{\mathrm{GeV}}$, and
  $m_{t}=174.3\,{\mathrm{GeV}}$ \cite{Perk}.}, respectively,
  \begin{equation}\label{eqn:DratesGs}
    \Gamma_{Z\rightarrow gg} = \left\{
      \begin{array}{c}
        5.10\cdot 10^{-7}\\
        9.60\cdot 10^{-15}
      \end{array}\right\}
      \biggl(\frac{B}{B_{0}}\biggr)^{2}\mathrm{MeV}.
  \end{equation}
  The values of the branching ratio of the above decay rate with the total
  width of the $Z$ in the absence of a magnetic field are
  \begin{equation}\label{eqn:BRGs}
    \frac{\Gamma_{Z\rightarrow gg}}{\Gamma}
      = \left\{
      \begin{array}{c}
        2.04\cdot 10^{-10}\\
        3.85\cdot 10^{-18}
      \end{array}\right\}
      \biggl(\frac{B}{B_{0}}\biggr)^{2}.
  \end{equation}

  For comparison purposes it is convenient to contrast the results
  of the two gluon case with the two photon case.  The results of
  the latter decay were determined in a manner completely
  analogous to that which has been outlined in this article for the
  decay $Z\rightarrow gg$.  The
  values of the decay rate for the process $Z\rightarrow
  \gamma\gamma$ are
  \begin{equation}\label{eqn:DratesPs}
    \Gamma_{Z\rightarrow \gamma\gamma} = \left\{
      \begin{array}{c}
        8.33\cdot 10^{-8}\\
        6.63\cdot 10^{-8}
      \end{array}\right\}
      \biggl(\frac{B}{B_{0}}\biggr)^{2}\mathrm{MeV}.
  \end{equation}
  Notice that the degree to which the two values differ is much smaller
  than in the two gluon
  case. This is because the major contribution to the process is due to the
  fermion mass being that of the electron, which is known very
  well.  The corresponding values for the branching ratio are
  \begin{equation}\label{eqn:BRPs}
    \frac{\Gamma_{Z\rightarrow \gamma\gamma}}{\Gamma}
      = \left\{
      \begin{array}{c}
        3.34\cdot 10^{-11}\\
        2.66\cdot 10^{-11}
      \end{array}\right\}
      \biggl(\frac{B}{B_{0}}\biggr)^{2}.
  \end{equation}

  \section{Conclusion}  This article has demonstrated that
  the existence of a magnetic field can stimulate decays that
  were otherwise forbidden.  Schwinger's powerful proper-time method
  was used in conjunction with a first order expansion in the
  field to determine
  what corrections exist for the Dirac Green's function.  These new
  contributions were analyzed in the specific decays
  $Z\rightarrow gg$ and $Z\rightarrow \gamma\gamma$.
  Though the resulting decay rates are small, they are non-zero.
  In addition it was also found that
  there exists a very large asymmetry as to whether the decay
  products will exit perpendicular or parallel to the magnetic field.

  \appendix
  \section{}
  As noted in the body of the article, the amplitude for the decay
  $Z\rightarrow gg$ to first order in an external magnetic field
  (Fig. \ref{fig:Zggs}) is the sum
  the contributions from the zeroth order amplitude (Eq.\ (\ref{eqn:Ma0})),
  first order corrections to each leg of the charged ferion loop
  (Eq.\ (\ref{eqn:Mapr1}) and the like), the first order correction to the phase
  of the amplitude (Eq.\ (\ref{eqn:Maph})), and similar corrections for the
  crossed process which one obtains by interchange of the labels (1,2).

  Evaluation of this amplitude gives the result found in
  Eq.\ (\ref{eqn:Amp11}).  Gauge invariance of this amplitude implies the
  following six relationships which we confirmed,
    \begin{subequations}\label{eqn:GIRelat}
    \begin{eqnarray}
      N_{6}&=&4N_{1}-4N_{2}-2N_{5}\\\label{eqn:GIR-6}
      N_{7}&=&2N_{2}\\\label{eqn:GIR-7}
      N_{8}&=&2N_{1}-2N_{2}\\\label{eqn:GIR-8}
      N_{9}&=&-2N_{1}\\\label{eqn:GIR-9}
      N_{10}&=&-2N_{3}\\\label{eqn:GIR-10}
      N_{11}&=&-2N_{4}\label{eqn:GIR-11}.
    \end{eqnarray}
    \end{subequations}
  Explicitly, the five remaining independent coefficients are
    \begin{subequations}\label{eqn:Coeff}
    \begin{eqnarray}
      N_{1}=\sum_{i}\frac{Q_{i}g_{v}^{i}}{\bigl(t_{i}-4\bigr)^{2}}&\times&
        \Bigl[\bigl(-128\,t_{i}^{2}+768\,t_{i}-1024\bigr)\nonumber\\
        & &+\bigl(96\,t_{i}^{2}-896\,t_{i}+2048\bigr)A(t_{i})\nonumber\\
        & &-\bigl(64\,t_{i}-512+1024\,t_{i}^{-1}\bigr)B(t_{i})\Bigr]
    \\\label{eqn:N1}
      N_{2}=\sum_{i}\frac{Q_{i}g_{v}^{i}}{\bigl(t_{i}-4\bigr)^{2}}&\times&
        \Bigl[\bigl(-8\,t_{i}^{3}+32\,t_{i}^{2}-256\,t_{i}+1024\bigr)
    \nonumber\\
       & &-\bigl(80\,t_{i}^{2}-448\,t_{i}+512\bigr)A(t_{i})\nonumber\\
       & &+\bigl(16\,t_{i}^{2}-64\,t_{i}-256+1024\,t_{i}^{-1}\bigr)B(t_{i})
    \Bigr]\\\label{eqn:N2}
      N_{3}=\sum_{i}\frac{Q_{i}g_{v}^{i}}{\bigl(t_{i}-4\bigr)^{2}}&\times&
        \Bigl[\bigl(-(8/3)\,t_{i}^{3}+(160/3)\,t_{i}^{2}-(512/3)\,t_{i}\bigr)
    \nonumber\\
       & &+\bigl(48\,t_{i}^{2}-320\,t_{i}+512\bigr)A(t_{i})\nonumber\\
       & &-\bigl(16\,t_{i}^{2}-128\,t_{i}+256\bigr)B(t_{i})\Bigr]
    \\\label{eqn:N3}
      N_{4}=\sum_{i}\frac{Q_{i}g_{v}^{i}}{\bigl(t_{i}-4\bigr)^{2}}&\times&
        \Bigl[\bigl(8\,t_{i}^{3}-64\,t_{i}^{2}+128\,t_{i}\bigr)\nonumber\\
       & &-\bigl(48\,t_{i}^{2}-384\,t_{i}+768\bigr)A(t_{i})\nonumber\\
       & &+\bigl(16\,t_{i}^{2}-128\,t_{i}+256\bigr)B(t_{i})\Bigr]
    \\\label{eqn:N4}
      N_{5}=\sum_{i}\frac{Q_{i}g_{v}^{i}}{\bigl(t_{i}-4\bigr)^{2}}&\times&
        \Bigl[\bigl(16\,t_{i}^{3}+64\,t_{i}^{2}-1024\,t_{i}+2048\bigr)
    \nonumber\\
       & &-\bigl(224\,t_{i}^{2}-1920\,t_{i}+4096\bigr)A(t_{i})\nonumber\\
       & &+\bigl(32\,t_{i}^{2}-128\,t_{i}-512
       +2048\,t_{i}^{-1}\bigr)B(t_{i})\Bigr],\label{eqn:N5}
    \end{eqnarray}
    \end{subequations}
  where $t_{i}$ is defined by the square of the ratio of the $Z$ mass to
  the quark mass
    \begin{equation}\label{eqn:ti}
      t_{i}=\biggl(\frac{M_{Z}}{m_{i}}\biggr)^{2},
    \end{equation}
  and the functions $A(t_{i})$ and $B(t_{i})$ are defined by \cite{Shima}
    \begin{equation}\label{eqn:At}
    \begin{array}{r c l @{\mathrm{\hspace{2mm}for\hspace{2mm}}} l}
      A(t)=\int_{0}^{1}\!\!\mathrm{d}x\ln[1-tx(1-x)-i\delta]&=&
    2\biggl[\sqrt{\frac{t-4}{t}}\sinh^{-1}\sqrt{\frac{-t}{4}}-1\biggr],
    & t\leq 0 \\[5mm]
       &=&2\biggl[\sqrt{\frac{4-t}{t}}\sin^{-1}\sqrt{\frac{t}{4}}-1\biggr],
    & 4\geq t\geq 0 \\[5mm]
       &=&2\biggl[\sqrt{\frac{t-4}{t}}
    \biggl(\cosh^{-1}\sqrt{\frac{t}{4}}-\frac{i\pi}{2}\biggr)-1\biggr],
    & t\geq 4,
    \end{array}
    \end{equation}
    and
    \begin{equation}\label{eqn:Bt}
    \begin{array}{r c l @{\mathrm{\hspace{2mm}for\hspace{2mm}}} l}
      B(t)=\int_{0}^{1}\!\!\mathrm{d}x\frac{\ln[1-tx(1-x)-i\delta]}{x(1-x)}&=&
    4\biggl[\sinh^{-1}\sqrt{\frac{-t}{4}}\biggr]^{2},
    & t\leq 0 \\[5mm]
       &=&-4\biggl[\sin^{-1}\sqrt{\frac{t}{4}}\biggr]^{2},
    & 4\geq t\geq 0 \\[5mm]
       &=&4\biggl[\cosh^{-1}\sqrt{\frac{t}{4}}\biggr]^{2}-\pi^{2}
    -4i\pi\cosh^{-1}\sqrt{\frac{t}{4}},
    & t\geq 4.\\
    \end{array}
    \end{equation}

  Notice that in Eqs.\ (\ref{eqn:Coeff}) there is a sum over all possible
  quarks in the charge loop.  For the decay $Z\rightarrow\gamma\gamma$ this
  could also include the electron.  $Q_{i}$ and $g_{v}^{i}$ are the charge
  and the associated vector coupling of the fermion, repectively.  The
  factors of $M_{Z}$ present are due to the
  evaluation of the dot product between the two gluon momenta
    \begin{eqnarray}\label{eqn:DotProd}
      k_{1}\cdot k_{2}&=&(k_{1} + k_{2})^{2}/2 \\\nonumber
       &=& k^{2}/2 \\\nonumber
       &=& M_{Z}^{2}/2,
    \end{eqnarray}
  where $k$ is the four-momentum of the $Z$.

  \begin{acknowledgments}
  I wish to extend special thanks to Duane Dicus for the useful
  discussions and help given throughout the course of this work.
  \end{acknowledgments}

\end{document}